\documentclass{article}
\usepackage{spconf,amsmath,amssymb,graphicx,hyperref,booktabs,multirow}
\usepackage[table]{xcolor}
\usepackage{siunitx}
\usepackage{tikz}
\usetikzlibrary{positioning,shapes.geometric, arrows}

\def\x{{\mathbf x}}
\def\s{{\mathbf s}}
\def\n{{\mathbf n}}
\newcommand{\shat}{\hat{\mathbf s}}
\newcommand{\nhat}{\hat{\mathbf n}}

\newcommand{\rowhl}{\rowcolor{gray!12}}

\title{SNOAR: Self-training with Noise filtering, Observation adding, and Remixing for Speech Denoising Without Clean References}

\twoauthors
  {Samuele Cornell}
    {Carnegie Mellon University, USA}
  {Matthew Maciejewski}
    {Johns Hopkins University, USA}

\begin{document}
\ninept
\maketitle

\begin{abstract}
Unsupervised speech denoising methods such as MixIT and noisy-target training can be trained directly on noisy recordings, avoiding the sim-to-real gap of synthetic training data, but remain well below supervised performance. Teacher--student self-training, as in RemixIT, narrows this gap by remixing a teacher's estimates into new mixtures that supervise a student; however, since the student is trained only toward the teacher's estimates, teacher errors, in particular over-suppressed speech, are never recovered and propagate as the teacher is updated from the student. We propose SNOAR, a self-training framework that corrects the teacher's estimates before use: the noise leaked into its speech estimate is removed by differential noise filtering (DNF), and part of the noisy observation is added back in to restore suppressed speech and attenuate artifacts. We show that the two corrections depend on each other, as without DNF the loop collapses to the identity mapping. Compared to RemixIT, SNOAR is more stable along training and outperforms it in signal-based, perceptual, and word error rate metrics, reaching 97\% of the SI-SDR improvement of a fully-supervised topline on WHAM!; on real CHiME-3 recordings, it improves every metric of three different teachers, while RemixIT degrades their perceptual quality. 
\end{abstract}


\section{Introduction}
\label{sec:intro}
Supervised speech denoising models are trained on pairs of noisy and clean speech, but
for real-world recordings no such pair exists: the clean signal is precisely
what the model is asked to produce. These systems are therefore trained
on synthetic mixtures, built by adding separately recorded noise to clean
speech captured under controlled conditions. This restricts training to
domains where studio-quality clean speech is available, which in practice often means read speech in high-resource languages. As a consequence, the inherent mismatch between such synthetic mixtures and real ``speech-in-the-wild'' recordings, the so-called ``sim-to-real gap'', is known to degrade performance at deployment~\cite{leglaive2023chime, leglaive2025objective}.

In recent years, in order to circumvent this issue, several methods have been proposed to learn directly from noisy recordings alone. Generative approaches leverage a clean-speech prior~\cite{bando2018vaenmf,bie2022unsupervised,nortier2024udiffse,jiang2023unse,xiang2020parallel} and thus still need an unpaired studio-quality clean speech corpus. Predictive approaches, on the other hand, such as noisy-target training (NyTT)~\cite{fujimura2021noisy, fujimura2023analysis}, mixture invariant training (MixIT)~\cite{wisdom2020unsupervised} and NyTT with differential noise filtering (DNF)~\cite{maciejewski2026exploiting}, need only in-domain noisy speech and noise-only clips. We argue that this approach is more scalable, as noise-only clips can be obtained from the same recordings by manual labeling and/or automatic voice activity detection~\cite{wisdom2020unsupervised}. 
Although these methods narrow sim-to-real generalization gap~\cite{leglaive2023chime}, 
they still fall well short of what we know we can currently achieve with supervised training on synthetic data, where perfectly clean references are available~\cite{maciejewski2026exploiting, tzinis2022remixit}.

Self-training pseudo-labeling frameworks such as RemixIT~\cite{tzinis2022continual, tzinis2022remixit} have been proposed to close more of this gap. 
RemixIT remixes a teacher model's speech and noise estimates across recordings into mixtures whose targets are now known pseudo-labels, trains a student on them, and periodically replaces the teacher with the student or updates it by an exponential moving average~\cite{tarvainen2017mean}. 
The teacher can be trained in any of three ways: supervised, on synthetic mixtures with clean references; unsupervised, on real recordings alone, as with MixIT; or jointly, on synthetic mixtures and real recordings together, as with DNF.
RemixIT was originally validated on synthetic mixtures with clean references~\cite{tzinis2022remixit, tzinis2022continual}. On real recordings, its record is less favorable: as the baseline of the CHiME-7 UDASE task~\cite{leglaive2023chime, leglaive2025objective}, its student reduced the background noise but ended below its own teacher in perceived speech and overall quality.
We argue that this follows from how RemixIT constructs its pseudo-labels: the student is trained toward the teacher's estimates, so whatever speech the teacher suppresses or distorts is reproduced by the student and, once the teacher is updated from it, passed on to the next generation. This is the regression analogue of confirmation bias in pseudo-labelling for classification~\cite{arazo2020pseudo, amini2025self}, where teacher errors are reinforced unless the pseudo-labels are corrected beforehand, which RemixIT never does.

Motivated by this, we propose self-training with noise filtering, observation adding and remixing (SNOAR), a self-training framework in which the teacher's errors are corrected before its estimates are used as self-supervision. A speech estimate departs from the clean speech in three ways: the noise it retains, the speech it suppresses, and the artifacts it introduces~\cite{iwamoto2022artifacts}, and SNOAR addresses them with two operations. The teacher's noise estimate is first rescaled and subtracted from its speech estimate, thus removing the retained noise residual using differential noise filtering. The result is then combined with the original noisy observation, which restores part of the suppressed speech and attenuates the artifacts, as in observation adding (OA)~\cite{iwamoto2022artifacts}. We show both experimentally and analytically that the two operations depend on each other: without DNF, OA drives the loop to collapse towards the identity mapping.

We compare SNOAR and RemixIT initialized from the same teachers with the same
training budget. On WHAM!~\cite{wichern19_interspeech}, SNOAR improves a MixIT teacher
from 8.3 to 13.3~dB SI-SDR~\cite{le2019sdr}, within 0.43~dB of a supervised
topline trained on the same mixtures, and obtains the lowest word error rate
(WER) and the highest perceptual scores of all unsupervised systems. Our analysis on WHAM! shows that both
methods remove the teacher's residual noise equally well and that the advantage of SNOAR comes from retaining more speech and introducing fewer artifacts; along training, RemixIT peaks early and then degrades, whereas SNOAR plateaus stably. On real CHiME-3
recordings~\cite{barker2015third}, SNOAR lowers the WER of three different teachers by 1.3 to 3.7 points and raises their UTMOS~\cite{saeki2022utmos} by 0.5 to 0.8, with the largest gain for the teacher trained without any clean
speech, whereas RemixIT degrades the perceptual scores of every teacher. Audio examples are available online\footnote{\href{https://popcornell.github.io/SNOAR-demo/}{popcornell.github.io/SNOAR-demo/}}.


\section{Method}
\label{sec:method}

\subsection{Speech denoising without clean references}
\label{sec:se_with_no_clean}
We consider a recording $\x=\s+\n_1$, where $\s$ is clean speech and $\n_1$
is real noise, neither observed separately, and a model $f_\theta$ that maps
a mixture to a speech estimate $\shat$ and a noise estimate $\nhat$.
NyTT~\cite{fujimura2021noisy, fujimura2023analysis}, MixIT~\cite{wisdom2020unsupervised} and
NyTT+DNF~\cite{maciejewski2026exploiting} train $f_\theta$ without clean references
by adding a second noise $\n_2$ from the same domain and supervising it on
$\x+\n_2$ with $\x$ and $\n_2$ as targets. As shown
in~\cite{maciejewski2026exploiting}, a network can separate speech from noise but
not $\n_1$ from $\n_2$, as the two come from the same distribution: the supervision asks for part of the inseparable noise mixture in one output and another part in the other, with the end result being that the trained network outputs the noise mixture in both outputs at some attenuation. For inference on $\x$, where no
noise is added, such a network outputs
\begin{equation}
  \shat = \s + \lambda\,\n_1 ,\qquad \nhat = \mu\,\n_1 ,
  \label{eq:lambda}
\end{equation}
for a residual fraction $\lambda$ and a noise gain $\mu$.

\newcommand{\sflt}{\shat^{\,\mathrm{dnf}}}
\newcommand{\soa}{\shat^{\,\mathrm{oa}}}

Differential noise filtering~\cite{maciejewski2026exploiting} removes this
residual using the model's own noise estimate: $\nhat$ is rescaled to the
magnitude of the residual in $\shat$ and subtracted from it,
\begin{equation}
  \sflt = \shat - \frac{\langle \nhat,\shat\rangle}{\|\nhat\|^2}\,\nhat ,
  \label{eq:proj}
\end{equation}
where $\sflt$ denotes the DNF-filtered speech estimate, which under
Eq.~\eqref{eq:lambda} is free of $\lambda\n_1$, since the two outputs carry
the same noise up to a scalar. The system evaluated in~\cite{maciejewski2026exploiting}, which trains with the NyTT objective and applies DNF at inference, is denoted NyTT+DNF in the rest of this paper. 

MixIT separates $\x+\n_2$ into three outputs, one for speech ($\shat$) and two for
noise ($\nhat_1$ and $\nhat_2$), and since only $\n_1+\n_2$ is separable from the speech, the two noise
outputs together carry the same noise as the single noise output above:
Eq.~\eqref{eq:lambda} holds also for MixIT with $\nhat$ being now their sum.
As in~\cite{tzinis2022remixit}, we train all our MixIT models with a mixture
consistency layer~\cite{wisdom2019differentiable}, so that the outputs sum to the
input and $\nhat=\x-\shat$. MixIT has no DNF-like operation at inference, so its
speech output retains whatever fraction $\lambda$ the converged network produces.

\subsection{SNOAR: self-training with noise filtering, observation adding and
remixing}
\label{sec:soar}

\textbf{RemixIT.} RemixIT~\cite{tzinis2022remixit} trains a student $f_\theta$
from the estimates of a teacher $f_{\bar\theta}$. For a batch of $B$
noisy speech recordings $\{\x_i\}$, with $\x_i=\s_i+\n_i$, the teacher produces
$(\shat_i,\nhat_i)$; the speech estimate of recording $i$ is remixed with the
noise estimate of a \emph{different} recording $j$, obtained by a circular
shift of the batch, into $\x'_i=\shat_i+\nhat_j$; and the student is trained
on $\x'_i$ toward the targets $(\shat_i,\nhat_j)$, which sum to its input. The
teacher is either replaced by the student every $K$ steps or tracks it using
an EMA of its weights.

\begin{figure*}[t]
\centering
\begin{tikzpicture}
\tikzset{font=\footnotesize}
\definecolor{blue}{RGB}{0, 45, 150}
\definecolor{green}{RGB}{0, 150, 0}
\definecolor{purple}{RGB}{128, 0, 128}

\tikzstyle{block} = [rectangle, rounded corners, inner sep=5pt, align=center, draw=black, line width=0.75pt]
\tikzstyle{netblock} = [rectangle, inner sep=5pt, align=center, draw=black, line width=1.5pt, rounded corners]
\tikzstyle{arrow} = [line width=0.75pt,->,>=stealth]

\node (x_i) [block, draw=black!50, fill=black!5] {noisy\\ recording\\ $\x_i$};
\node (ema) [netblock, right=10pt of x_i, draw=blue!75, fill=blue!15] {EMA\\teacher\\$f_{\bar{\theta}}$};
\node (dnf) [block, right=13pt of ema, draw=yellow!50!black, fill=yellow!15] {\textbf{Differential noise filtering}\\$\shat_i^\text{dnf} = \shat_i - \tfrac{\langle\nhat_i,\shat_i\rangle}{||\nhat_i||^2}\nhat_i$};
\node (oa) [block, right=18pt of dnf, yshift=4pt, draw=green!60!black, fill=green!15] {\textbf{Observation adding}\\$\shat_i^\text{oa} = (1-\gamma)\shat_i^\text{dnf}+\gamma\x_i$};
\node (remix) [block, inner ysep=7pt, right=17pt of oa, yshift=-8pt, draw=purple!80!black, fill=purple!15] {\textbf{Remixing}\\$\x_i'=\shat_i^\text{oa}+\nhat_j$};
\node (student) [netblock, right=13pt of remix, draw=blue!75, fill=blue!15] {EMA\\student\\$f_\theta$};
\node (loss) [block, right=18pt of student, draw=red!60!black, fill=red!15] {\textbf{SI-SDR loss}\\$\ell(\shat_i^\text{stu}, \shat_i^\text{oa})$\\$+\ell(\nhat_i^\text{stu},\nhat_j)$};

\node (x_j) [block, dashed, draw=black!50, fill=black!5, below=7pt of dnf, xshift=45pt] {other recording \\ $\x_j$};
\node (ema_copy) [netblock, dashed, right=10pt of x_j, draw=blue!75, fill=blue!15] {EMA\\ teacher\\{\color{black!60}(same weights)}};

\draw [arrow] (x_i) -- (ema);
\draw [arrow] ([yshift=8pt]ema.east) -- ([yshift=8pt]dnf.west) node[midway,below] {$\shat_i$};
\draw [arrow] ([yshift=-8pt]ema.east) -- ([yshift=-8pt]dnf.west) node[midway,below] {$\nhat_i$};
\draw [arrow] (dnf) -- ([yshift=-4pt]oa.west) node[midway,below] {$\shat_i^\text{dnf}$};
\draw [arrow] (oa) -- ([yshift=8pt]remix.west) node[midway,below] {$\shat_i^\text{oa}$};
\draw [arrow] (remix) -- (student) node[midway,below] {$\x_i'$};
\draw [arrow] ([yshift=8pt]student.east) -- ([yshift=8pt]loss.west) node[midway,below] {$\shat_i^\text{stu}$};
\draw [arrow] ([yshift=-8pt]student.east) -- ([yshift=-8pt]loss.west) node[midway,below] {$\nhat_i^\text{stu}$};
\draw [arrow,dashed,color=black!50] (x_j) -- (ema_copy);

\draw [arrow] ([xshift=4pt]x_i.east) -- ([xshift=4pt,yshift=23.8pt]x_i.east) -| ([xshift=-11,yshift=4pt]oa.west) -- ([yshift=4pt]oa.west);

\draw [arrow] ([yshift=8pt,xshift=-8.5pt]remix.west) -- ([yshift=26pt,xshift=-8.5pt]remix.west) -| (loss);
\draw [arrow,dashed,color=black!50] ([yshift=8pt]ema_copy.east) -- ([yshift=8pt,xshift=15pt]ema_copy.east) node[midway,below] {$\shat_j$};
\draw [arrow] ([yshift=-8pt]ema_copy.east) node[below,xshift=7.5pt] {$\nhat_j$} -| (loss.south) ;
\draw [arrow] ([yshift=-8pt,xshift=25pt]ema_copy.east) |- ([yshift=-8pt]remix.west);

\draw [arrow,dashed,draw=blue!75] ([xshift=-5pt]student.south) |- ([yshift=-48pt]ema.south) node[right,yshift=6pt]{{\color{blue!75}EMA update:} $\bar{\theta}\leftarrow\eta\bar{\theta}+(1-\eta)\theta$} -- (ema);
\draw [arrow,dashed,draw=red!60!black] ([xshift=10]loss.south) |- ([xshift=5pt,yshift=-44.3pt]student.south) node[right,yshift=6pt]{\color{red!60!black}Backpropagation} -- ([xshift=5pt]student.south);
\end{tikzpicture}
\vspace{-5pt}
\vspace{-10pt}
\caption{The SNOAR self-training framework. From a real recording $\x_i$ the
EMA teacher produces a speech and a noise estimate; the speech estimate is
filtered against the noise estimate by DNF (Eq.~\ref{eq:proj}) and combined
with the observation by OA (Eq.~\ref{eq:oa}), giving the corrected
self-supervision $\soa_i$. This is remixed with the noise estimate of another
recording $\x_j$ to form the student input. The student is trained to
reconstruct $\soa_i$ and $\nhat_j$, and the teacher is updated from the student weights by EMA.}
\label{fig:block}
\vspace{-12pt}
\end{figure*}
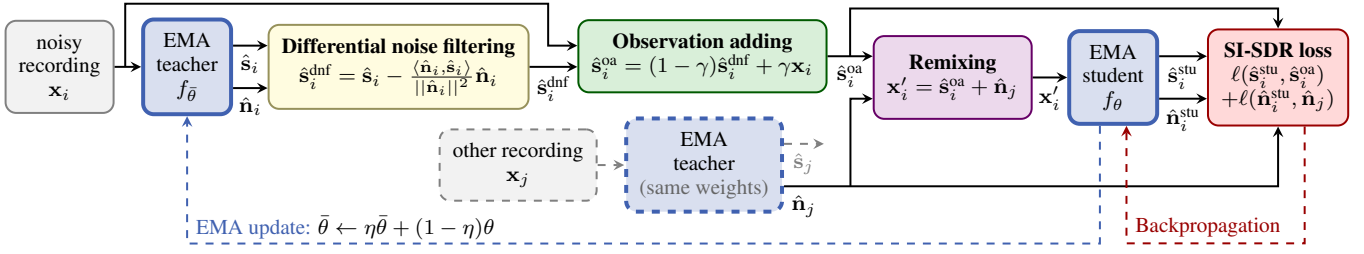

\textbf{Self-supervision correction.} SNOAR keeps the remixing step and the
loss of RemixIT and replaces the speech target by a corrected one, in two steps
(Fig.~\ref{fig:block}). First, the teacher's speech estimate is filtered
against the teacher's own noise estimate by DNF: Eq.~\eqref{eq:proj} is
applied to $(\shat_i,\nhat_i)$, giving $\sflt_i$. For a NyTT+DNF teacher this
is already its inference rule; for a MixIT teacher $\nhat_i$ is the sum of the
two noise outputs. In both cases, under Eq.~\eqref{eq:lambda}, this removes
the leaked $\lambda\n_i$ from the self-supervision (at the cost of increased artifacts). Second, the filtered estimate is
combined with the original noisy speech observation,
\begin{equation}
  \soa_i = (1-\gamma)\,\sflt_i + \gamma\,\x_i ,
  \label{eq:oa}
\end{equation}
referred to as observation adding (OA)~\cite{iwamoto2022artifacts}, originally proposed as an inference-time post-processing technique to improve speech recognition.
We instead apply it here to the self-supervision. Any speech estimate can be decomposed as
$\shat=\alpha\s+\lambda\n+\mathbf a$, with $\alpha\s$ the retained speech,
$\lambda\n$ the residual noise, and $\mathbf a$ the component orthogonal to
the subspace spanned by $\s$ and $\n$, i.e.\ the nonlinear distortion
introduced by the model itself~\cite{vincent2006performance}. Unlike the residual
noise, $\mathbf a$ has no counterpart in the input, so the only way to reduce
it is to add back a part of the input itself. Since $\x_i=\s_i+\n_i$ contains
all of the speech and no distortion, Eq.~\eqref{eq:oa} attenuates
$\mathbf a_i$ and $1-\alpha_i$ by $1-\gamma$, at the cost of returning a
fraction $\gamma$ of the recording's noise to the self-supervision. 
In vanilla RemixIT, the self-supervision is the teacher's estimate itself, so $\alpha_i$ and $\mathbf a_i$ pass to the student unchanged and propagate across training, which makes RemixIT susceptible to the confirmation bias problem discussed in Sec.~\ref{sec:intro}.

\textbf{OA noise propagation.}
One of the challenges with teacher-student training is the propagation of behaviors: if the teacher model starts to do something bad, the student learns from it, which then becomes the teacher, so the behavior propagates and can even be exacerbated.
This behavior could be observed with $\lambda$, the scaling of the noise residual.
Assuming mixture consistency (i.e. $\mu = 1-\lambda$) and that the noise recordings have approximately the same energy (i.e. $||\n_1||^2 \approx ||\n_2||^2$), when the student model is trained with the teacher outputs $\s_i+\lambda\n_i$ and $(1-\lambda)\n_j$, the optimal \textit{new} $\lambda'$ is:
\begin{equation}
\lambda' = \frac{\lambda^2}{\lambda^2+(1-\lambda)^2} \label{eq:lambda_update}\text{,}
\end{equation}
i.e. the Wiener gain for $\lambda\n_i$ in the presence of $(1-\lambda)\n_j$.

If we consider the equilibrium points of Eq.~\eqref{eq:lambda_update}, where $\lambda' = \lambda$, there are three points: $0$, $0.5$, and $1$.
Both $0$ and $1$ are stable, while $0.5$ is not, meaning that for $\lambda \in [0, 0.5)$, over iterations it will converge to $0$, and for $\lambda \in (0.5, 1]$, it will converge to $1$.
So, it is critical to keep $\lambda$ under $0.5$.

With the addition of OA, the equation instead becomes:
\begin{equation}
\lambda' = \frac{(\lambda+\gamma)^2}{(\lambda+\gamma)^2+(1-\lambda)^2} \label{eq:lambda_update_oa}\text{,}
\end{equation}
which shifts the $0.5$ transition point to $\tfrac{1}{4}(1-2\gamma+\sqrt{1-4\gamma-4\gamma^2})$.
For example, for $\gamma=0.2$, this requires $\lambda$ to be less than $0.2$, which corresponds to a noise reduction of at least $\approx14$~dB by the teacher model.
As a result, there is a large risk of the model collapsing to $\lambda=1$, i.e. no denoising and simply outputting the noisy speech.

With the introduction of DNF, since the residual noise is canceled, $\lambda'$ no longer scales from previous values of $\lambda$, only the OA parameter:
\begin{equation}
\lambda' = \frac{\gamma^2}{\gamma^2+(1-\lambda)^2} \label{eq:lambda_update_dnf_oa}\text{,}
\end{equation}
which has a transition point at $\lambda=\frac{1}{2}(1+\sqrt{1-4\gamma^2})$ and stable equilibria at $\lambda=\frac{1}{2}(1-\sqrt{1-4\gamma^2})$ and $\lambda=1$.
For $\gamma=0.2$, the transition point is at $\approx0.958$ and lower equilibrium is at $\approx0.042$.
This is almost certain to converge to the lower value, and although it is not $0$, it would correspond to a noise reduction of $\approx27.6$~dB, which is far more than any of our teachers achieves.


 
\textbf{Training objective and teacher update.} The student input is $\x'_i=\soa_i+\nhat_j$,
its targets are $\soa_i$ and $\nhat_j$, and it is trained with the
negative SI-SDR~\cite{le2019sdr} on both heads,
$\mathcal{L}=\ell(\shat^{\text{stu}}_i,\soa_i)+
\ell(\nhat^{\text{stu}}_i,\nhat_j)$, followed by the EMA update
$\bar\theta\leftarrow\eta\bar\theta+(1-\eta)\theta$. 
At inference, the student is used as its teacher was, i.e., with
Eq.~\eqref{eq:proj} for a two-output (speech + noise) model, or by taking the
speech output alone for a three-output MixIT-derived model. Both SNOAR and
RemixIT are agnostic to how the teacher was trained: fully supervised on
synthetic mixtures, fully unsupervised (NyTT, MixIT, NyTT+DNF), or a hybrid of
the two.
 
\vspace{-14pt}

\section{Experimental setup}
\label{sec:setup}

\textbf{Data.} We train on WHAM!~\cite{wichern19_interspeech} at 16~kHz, the official anechoic single-speaker mixtures of clean speech and real ambient noise, using the \texttt{tr} split. To simulate a real world application, the noise clips of \texttt{tr} are divided into two disjoint pools, one providing the noise $\n_1$ inside the recordings and the other the added noise $\n_2$, so that no clip is ever both the unobserved noise of one example and a known target of another. We test on \texttt{tt}, both microphones, on the untruncated \texttt{max} version. As such, results are not directly comparable with~\cite{maciejewski2026exploiting}. 

We also use CHiME-3~\cite{barker2015third}, which provides real recordings with no clean reference: we adapt on \texttt{tr05\_real} (channel~5), select checkpoints on the real development set \texttt{dt05\_real}, and test on \texttt{et05\_real}. The added noise $\n_2$ is drawn from the official background pool used to build the simulated partitions (\texttt{tr05\_simu}), which contains no speech and was recorded in sessions separate from the speech utterances.

\textbf{Model and training.} We use TF-GridNet~\cite{wang2023tf} as
the architecture for every system, in the configuration of \cite{maciejewski2026exploiting}
(512-point short-time Fourier transform, hop size 128, 4 blocks, 24
channels, 192 hidden units, 4.0~M parameters). Every system is trained with negative SI-SDR~\cite{le2019sdr} loss, and MixIT-based systems additionally use a mixture-consistency layer~\cite{wisdom2019differentiable}. On WHAM!, teachers are trained for 150k steps with Adam, a learning rate of $10^{-3}$, batch size 8 and $\ell_2$ gradient-norm clipping at 5, saving a checkpoint every 10k steps, and self-training SNOAR and RemixIT runs are warm-started from the teacher and run for 150k steps with $\gamma=0.2$ and a checkpoint every 10k steps. Both use an EMA teacher, updated every step with decay $0.999$. On CHiME-3, since the amount of data is less, self-training runs instead for 7.5k steps with a checkpoint every 1000 steps. We do not use early-stopping: each is trained to its full budget and the reported checkpoint is chosen afterwards among those saved using UTMOS (on \texttt{cv} for WHAM! and \texttt{dt05\_real} for CHiME-3) which is reference free. Hyperparameters were tuned using UTMOS on each dataset respective development set.
The only exception is the WHAM! supervised topline, whose checkpoint and hyperparameters are chosen by oracle SI-SDR.


\textbf{Evaluation.} All outputs are normalised to $-26$~dBov (decibel-to-overload signal level) before scoring, since DNSMOS and UTMOS are sensitive to the level and the output scale. We report SI-SDR~\cite{le2019sdr}, STOI~\cite{taal2011algorithm}, wideband PESQ~\cite{rix2001perceptual}, DNSMOS P.835~\cite{reddy2021dnsmos} (D-SIG, D-BAK and D-OVRL), UTMOS~\cite{saeki2022utmos}, and the WER of the OWSM v4~\cite{peng2025owsm} base model, computed with VERSA~\cite{shi2025versa}.

\vspace{-10pt}
\section{Experimental Results}
\label{sec:exp_results}

\begin{table}[t]
\centering
\caption{Results on the WHAM! test set. SNOAR and RemixIT are listed under the teacher they were initialised from (MixIT or DNF). For SI-SDR, the 95\% confidence interval obtained via paired bootstrap over the test utterances has a half-width of at most $0.08$~dB for every system.}
\label{tab:wham}
\setlength{\tabcolsep}{1.6pt}
\footnotesize
\sisetup{
    reset-text-series = false,
    text-series-to-math = true,
    mode=text,
    round-mode=places,
    round-precision=2,
    table-number-alignment=center,
    detect-weight=true}
\begin{tabular}{@{}lSS[table-format=1.3,round-precision=3]SSSS[table-format=2.1,round-precision=1]@{}}
\toprule
System & {SI-SDR} & {STOI} & {PESQ} & {D-OVRL} & {UTMOS} & {WER} \\
\midrule
Unprocessed        & -1.48 & 0.751 & 1.10 & 1.46 & 1.38 & 19.8 \\
Clean              & {$\infty$} & 1.000 & 4.64 & 3.33 & 4.22 & \phantom{0}4.7 \\
Supervised topline & 13.68 & 0.952 & 2.26 & 3.21 & 3.60 & \phantom{0}9.5 \\
\midrule
NyTT~\cite{fujimura2021noisy} & \phantom{0}0.20 & 0.776 & 1.13 & 1.60 & 1.36 & 18.1 \\
\midrule
NyTT+DNF~\cite{maciejewski2026exploiting}, teacher & \phantom{0}4.87 & 0.847 & 1.32 & 2.43 & 2.05 & 17.7 \\
\ \ RemixIT(NyTT+DNF)   & \phantom{0}9.32 & 0.913 & 1.56 & 2.61 & 2.67 & 13.1 \\
\ \ SNOAR(NyTT+DNF) & 10.03 & 0.932 & 1.77 & 3.05 & 3.09 & 12.6 \\
\midrule
MixIT~\cite{wisdom2020unsupervised}, teacher & \phantom{0}8.32 & 0.872 & 1.54 & 2.54 & 2.26 & 15.3 \\
\ \ RemixIT(MixIT) & 12.10 & 0.926 & 1.61 & 2.98 & 3.15 & 13.3 \\
\rowhl \ \ SNOAR(MixIT) & \bfseries 13.25 & \bfseries 0.942 & \bfseries 2.02 & \bfseries 3.15 & \bfseries 3.45  & \bfseries 10.9 \\
\bottomrule
\end{tabular}
\end{table}

\begin{figure}[t]
\centering
\includegraphics[width=\columnwidth]{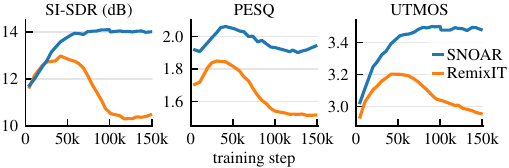}
\vspace{-18pt}
\caption{SNOAR(MixIT) and RemixIT(MixIT) at every training checkpoint, scored on a randomly selected subset of 500 WHAM! \texttt{cv} utterances.}
\label{fig:curves}
\vspace{-14pt}
\end{figure}
\vspace{-4pt}
\subsection{Results on WHAM!}
\label{sec:wham}

Table~\ref{tab:wham} reports both self-training methods from the same two
teachers. From the MixIT teacher, SNOAR reaches 13.25~dB SI-SDR against
12.10~dB for RemixIT, i.e.\ within 0.43~dB of the supervised topline trained
on the same mixtures with clean targets, and it attains the lowest WER of all
systems trained without clean references (10.9\%, against 9.5\% for the
topline). Its UTMOS, 3.45, is likewise within 0.15 of the topline's 3.60,
against 2.26 for the teacher it started from. The ordering is the same from
the weaker NyTT+DNF teacher, with a smaller margin. RemixIT improves every
metric over both teachers, but its gains in PESQ and WER are markedly smaller
than those of SNOAR
while the two methods are much closer in SI-SDR. RemixIT thus removes noise about as well as SNOAR does, but at a higher cost to the speech itself.

\label{sec:mech}
Table~\ref{tab:abl} reports the decomposition of Sec.~\ref{sec:method} for
the MixIT teacher and its two students, and confirms this interpretation.
We report the least-squares fit $\shat\approx\alpha\s+\beta\n$ on each test
utterance, with $\lambda=\beta/\alpha$ the noise coefficient relative to the
speech one and $r=\|\shat-\alpha\s-\beta\n\|^2/\|\shat\|^2$ the artifact
energy as a fraction of the estimate's, averaged over utterances. Both
students reduce the teacher's residual noise from $\lambda=0.218$ to about
$0.03$, as the fixed points of Sec.~\ref{sec:soar} predict: from this teacher, Eq.~\eqref{eq:lambda_update} converges to $0$ for RemixIT and
Eq.~\eqref{eq:lambda_update_dnf_oa} to $0.042$ for SNOAR with $\gamma=0.2$.
Remixing thus filters the noise of the self-supervision whether it was
leaked by the teacher or reintroduced by OA. The speech and artifact terms
behave differently. RemixIT lowers the artifact fraction from the teacher's
$r=0.085$ to $0.062$, whereas SNOAR reaches $0.048$, and it retains more of
the speech ($\alpha=0.949$ against $0.939$). The gain of SNOAR over RemixIT
is thus not obtained by removing more noise, but by preserving more
speech and introducing less distortion. The two methods also differ along
training (Fig.~\ref{fig:curves}): RemixIT peaks after a few tens of
thousands of steps and degrades thereafter on all three metrics, so that the
results in Table~\ref{tab:wham} are only attained by selecting its checkpoint,
whereas SNOAR generally increase and plateaus.

The bottom block of Table~\ref{tab:abl} ablates the two corrections and the
choice of $\gamma$. The $\gamma=0.2$ self-supervision carries a fifth of the
recording's noise and the $\gamma=0$ one carries none, yet the two students
end at $\lambda=0.031$ and $0.018$, against the $0.042$ and $0$ predicted by
Eq.~\eqref{eq:lambda_update_dnf_oa}.
OA acts instead on the remaining
two terms, raising $\alpha$ from $0.944$ to $0.949$ and lowering $r$ from
$0.059$ to $0.048$, which is worth 0.9~dB SI-SDR and 0.3 WER points. With
DNF alone ($\gamma=0$) the student ends close to the selected RemixIT
checkpoint on every metric (12.33 against 12.10~dB, $r=0.059$ against
$0.062$), so most of the gain of SNOAR over RemixIT is due to OA, while DNF
is what makes OA usable: without it, Eq.~\eqref{eq:lambda_update_oa} leaves
the teacher at the edge of the basin of its non-collapsed fixed point, and
the student indeed ends at $\lambda=0.697$ and $\alpha=0.979$, i.e.\
reproducing its input, with the highest artifact fraction of the table.
Annealing $\gamma$ from 0.2 to 0.05 yields the same $\alpha$, $\lambda$ and
$r$ as the fixed value, so we keep $\gamma$ fixed. Finally, $\gamma=0.5$
diverges, as Eq.~\eqref{eq:lambda_update_dnf_oa} predicts: for
$\gamma\ge0.5$ its only stable fixed point is $\lambda=1$.

\begin{table}[t]
\centering
\caption{Ablation of SNOAR(MixIT) on the WHAM! \texttt{tt} set. The first row
is the method as described ($\gamma=0.2$). $\alpha$, $\lambda$ and $r$ are the
speech, noise and artifact terms of the decomposition of Sec.~\ref{sec:method}, with $r$ the
artifact energy as a fraction of the estimate's.}
\label{tab:abl}
\setlength{\tabcolsep}{1.6pt}
\footnotesize
\sisetup{
    reset-text-series = false,
    text-series-to-math = true,
    mode=text,
    round-mode=places,
    round-precision=2,
    table-number-alignment=center,
    detect-weight=true}
\begin{tabular}{@{}l@{\hspace{\tabcolsep}}SSSS[table-format=2.1,round-precision=1]S[table-format=1.3,round-precision=3]S[table-format=1.3,round-precision=3]S[table-format=1.3,round-precision=3]@{}}
\toprule
Arm & {SI-SDR} & {PESQ} & {UTMOS} & {WER} & {$\alpha$} & {$\lambda$} & {$r$} \\
\midrule
MixIT teacher & \phantom{0}8.32 & 1.54 & 2.26 & 15.3 & 0.912 & 0.218 & 0.085 \\
\rowhl SNOAR(MixIT), $\gamma{=}0.2$ & \bfseries 13.25 & \bfseries 2.02 & 3.45 & \bfseries 10.9 & 0.949 & 0.031 & 0.048 \\
RemixIT(MixIT) & 12.10 & 1.61 & 3.15 & 13.3 & 0.939 & 0.026 & 0.062 \\
\midrule
\ \ $\gamma{=}0.5$ & -0.82 & 1.11 & 1.38 & 19.9 & 0.998 & 0.899 & 0.033 \\
\ \ $\gamma{=}0.1$ & 13.14 & 1.87 & \bfseries 3.47 & 11.1 & 0.946 & 0.020 & 0.050 \\
\ \ $\gamma{=}0$ (no OA) & 12.33 & 1.84 & 3.27 & 11.2 & 0.944 & 0.018 & 0.059 \\
\ \ $\gamma$ annealed $0.2\!\to\!0.05$ & 13.23 & 1.84 & 3.43 & 11.6 & 0.945 & 0.020 & 0.048 \\
\ \ no DNF, Eq.~\eqref{eq:proj} & 0.23 & 1.13 & 1.38 & 17.6 & 0.979 & 0.697 & 0.135 \\
\bottomrule
\end{tabular}
\vspace{-12pt}
\end{table}

\begin{table}[h]
\centering
\caption{Results on CHiME-3 \texttt{et05\_real}, channel 5. No clean reference exists for these recordings, so only reference-free scores are reported. Each block is one in-domain teacher, trained without clean in-domain speech but allowed out-of-domain clean speech from WHAM! or LibriSpeech, followed by SNOAR and RemixIT adapted from it with the same budget of 7.5k steps.}
\label{tab:chime}
\setlength{\tabcolsep}{2.5pt}
\footnotesize
\sisetup{
    reset-text-series = false,
    text-series-to-math = true,
    mode=text,
    round-mode=places,
    round-precision=2,
    table-number-alignment=center,
    detect-weight=true}
\begin{tabular}{@{}l@{\hskip 4pt}lSSSSS[table-format=2.1,round-precision=1]@{}}
\toprule
Teacher & System & {D-SIG} & {D-BAK} & {D-OVRL} & {UTMOS} & {WER} \\
\midrule
 & Unprocessed & 2.05 & 1.50 & 1.48 & 1.38 & 22.0 \\
\midrule
\multirow{3}{*}{MixIT\,(real)}
 & Teacher & 2.77 & 2.01 & 1.94 & 1.62 & 22.1 \\
 & RemixIT & 2.67 & 1.95 & 1.87 & 1.57 & 20.3 \\
\rowhl & SNOAR & \bfseries 3.09 & \bfseries 2.37 & \bfseries 2.19 & \bfseries 2.10 & \bfseries 18.4 \\
\midrule
\multirow{3}{*}{Mix+sup\,(WSJ)}
 & Teacher & 3.13 & 2.46 & 2.28 & 2.03 & 17.1 \\
 & RemixIT & 2.70 & 2.01 & 1.91 & 1.69 & 18.4 \\
\rowhl & SNOAR & \bfseries 3.41 & \bfseries 3.14 & \bfseries 2.68 & \bfseries 2.75 & \bfseries 15.4 \\
\midrule
\multirow{3}{*}{Mix+sup\,(Libri)}
 & Teacher & 3.06 & 2.33 & 2.19 & 1.87 & 17.9 \\
 & RemixIT & 2.81 & 2.09 & 1.98 & 1.74 & 17.5 \\
\rowhl & SNOAR & \bfseries 3.35 & \bfseries 2.89 & \bfseries 2.53 & \bfseries 2.63 & \bfseries 16.6 \\
\bottomrule
\end{tabular}
\vspace{-12pt}
\end{table}

\vspace{-12pt}
\subsection{Results on CHiME-3 real noisy speech}
\label{sec:chime}

Table~\ref{tab:chime} repeats the comparison on real recordings, for three
teachers differing in the clean speech they were allowed. The first is trained
with MixIT on CHiME-3 real data only and uses no clean speech at all. The
other two are trained jointly on the same real data and on synthetic mixtures
built from out-of-domain studio recordings, either the WSJ~\cite{wsj0} speech of WHAM! or
LibriSpeech read audiobooks. The former matches the speech domain of CHiME-3,
whose utterances are also read WSJ excerpts, and is the strongest of the three
teachers on every metric.

SNOAR improves every metric of every teacher, reducing WER by 3.7, 1.7 and 1.3
points and raising UTMOS by 0.48, 0.72 and 0.76. Its largest gain is on the
teacher trained without any clean speech, which is the setting the method
targets. RemixIT falls below its teacher on every reference-free metric for all three, and lowers WER only for the two weaker ones; replacing the EMA
update with the periodic teacher replacement of~\cite{tzinis2022remixit}
did not help. This matches the CHiME-7 UDASE
baseline~\cite{leglaive2023chime, leglaive2025objective}, whose RemixIT student reduced the background noise but ended below its teacher in perceived quality, and the WHAM! decomposition of Sec.~\ref{sec:wham}, where RemixIT removed the noise
as well as SNOAR but retained less speech and more distortion, to which the reference-free metrics are sensitive. On CHiME-3 the teachers are weaker to begin with, since the recordings are noisier and the noise more diverse than on WHAM!, so their errors are larger and the confirmation bias of RemixIT is enough to cancel the benefit of self-training: the speech the teacher suppresses is absent from its self-supervision and each generation inherits the loss, whereas SNOAR corrects the self-supervision before use.

\vspace{-8pt}
\section{Conclusion}
\label{sec:conclusion}

We introduced SNOAR, a self-training framework for speech denoising without
clean references in which the teacher's estimates are corrected before they
are used as self-supervision: DNF removes the noise the teacher leaks, and OA
restores part of the speech it suppresses while attenuating its artifacts. 
We showed that the two corrections are not independent, by analysing the
equilibrium points of $\lambda$, the noise left in the speech estimate,
across self-training generations. Without DNF, the student's $\lambda$
depends on the teacher's, and OA lowers the threshold above which the loop
collapses to the identity mapping; with DNF, it depends only on the OA
coefficient $\gamma$ and the loop converges from any teacher to a residual
of $\approx27$~dB below the input noise, far below what any teacher achieves on its own. 
On WHAM!, SNOAR comes within 0.43~dB of
a supervised topline starting from a MixIT teacher 5~dB below it, and on
real CHiME-3 recordings it improves WER and reference-free metric of three
different teachers, whereas RemixIT falls below each of its teachers on all
of them. 


\vfill\pagebreak
\bibliographystyle{IEEEbib}

\bibliography{refs}
\end{document}